\documentclass[usenatbib]{mn2e}
\usepackage{epsfig}
\usepackage{pslatex}
\usepackage{amsmath}
\usepackage{fleqn}
\usepackage{breqn}

\title[Period Detection Based on Serial Dependence]{
Detection of Periodicity Based on Serial Dependence of Phase-Folded Data}
\author[Shay Zucker]{Shay Zucker\thanks{E-mail:
shayz@post.tau.ac.il}\\ Dept. of Geosciences, Raymond and Beverly
Sackler Faculty of Exact Sciences, Tel-Aviv University, Tel Aviv
6997801, Israel}
\begin{document}

\maketitle

\begin{abstract}	

We introduce and test several novel approaches for periodicity
detection in unevenly-spaced sparse datasets. Specifically, we examine
five different kinds of periodicity metrics, which are based on
non-parametric measures of serial dependence of the phase-folded
data. We test the metrics through simulations in which we assess their
performance in various situations, including various periodic signal
shapes, different numbers of data points and different signal to noise
ratios.  One of the periodicity metrics we introduce seems to perform
significantly better than the classical ones in some settings of
interest to astronomers.  We suggest that this periodicity metric --
the Hoeffding-test periodicity metric -- should be used in addition to
the traditional methods, to increase periodicity detection
probability.

\end{abstract}


\section{Introduction}

Detecting periodicity in an unevenly-sampled time series is a task one
frequently faces in many fields of astronomy. Astronomers who study
variable star light curves are probably the ones who face this
challenge most often, but it is also common in the analysis of radial
velocities (RV) of spectroscopic binary stars. The field of
time-domain astronomy is becoming increasingly important. Large
Time-domain surveys are already running (e.g. PTF \citep{Lawetal2009},
CRTF \citep{Draetal2012}, Pan-STARRS \citep{Kaietal2010}) or planned
(e.g. LSST \citep{LSS2009}), and some space-based astronomical
missions are basically time-domain surveys (e.g. {\it Kepler}
\citep{Kocetal2010}, {\it CoRoT} \citep{Auvetal2009}, {\it Hipparcos}
\citep{ESA1997}, {\it Gaia} \citep{Jor2008}). The data analysis
challenges that accompany the emerging huge databases emphasize the
importance of periodicity detection in unevenly-sampled time series.

During the years many researchers proposed methods and algorithms to
tackle the problem. A common approach is to calculate some kind of a
periodicity metric function, which provides a periodicity score for
each trial period. Period detection then consists of identifying a
peak in the periodicity metric function that is significantly higher
than all the other values. Depending on the specific kind of metric,
one may sometimes look for a trough instead of a peak. In any case,
the value of the periodicity score should be extreme at the correct
value of the period, compared to its value at other periods.

An inherent difficulty in periodicity analysis of astronomical time
series is the uneven time sampling. In classical time series analysis,
Fourier techniques are available that decompose the original signal
into a superposition of sinusoids. This approach is based on the
orthogonality properties of evenly-sampled sinusoids. This cannot be
done when the data are unevenly sampled, which is the common case in
astronomy, since the sinusoids are no longer orthogonal.

Some methods devised for the unevenly-sampled case try to fit to the
data some kind of a periodic function. The periodicity score is
usually related in one way or another to the $\chi^2$ statistic of the
fit. This is the 'least-squares' group of techniques. The techniques
in this group differ by the details of the periodic function and the
exact calculation of the periodicity score. The most commonly used
technique in this group is the Lomb--Scargle periodogram
\citep{Lom1976,Sca1982}.  Inspired by Fourier analysis, in
Lomb--Scargle periodogram one fits a sinusoid to the time series.
Another least-squares technique is the AoV method, which basically
fits a periodic piecewise constant function \citep{Sch1989}. Other
techniques exist that are based on the same general idea. Those
methods are very powerful if the actual shape of the periodic signal
is indeed close to the function the method assumes.

Another group of techniques is the 'string-length' group. The
classical methods in this group measure the sum of the squares of the
differences between one data point and the next, after the points have
been ordered in phase for a given trial period. \citet{Cla2002}
provides an overview of the string-length techniques, focusing on the
Lafler--Kinman method \citep{LafKin1965}, and the methods of
\citet{Ren1978} and \citet{Dwo1983}. In all the string-length methods,
one has to actually phase-fold the measurements for each trial period,
and then quantify the serial dependence of the measurements, i.e., the
dependence between consecutive measurements. The various string-length
statistics used in those methods can all be traced back to the
von-Neumann ratio statistic, also known as the Abbe value, or the
Durbin--Watson statistic, which was used originally to detect
autocorrelation in lag one in evenly-sampled time series
\citep{vonetal1941,DurWat1950,DurWat1951}:
\begin{equation}
\eta = \frac {\sum (x_{i+1}-x_i)^2} {\sum x_i^2}
\end{equation}
The advantage of string-length techniques is obvious when we do not
know in advance the shape of the periodic function we seek. The
underlying assumption is that phase-folding the data at the correct
period will produce a signal with some regularity, which will reflect
as serial dependence of the data points.

In the current paper, we propose to take the string-length philosophy
one step further and examine various non-parametric tests for serial
dependence between consecutive phase-folded measurements. Although the
term 'non-parametric' is used extensively in statistics, its
definition remains somewhat blurred. The basic idea is using as few
assumptions as possible about the probability distribution of the
quantity we study. A common feature to many non-parametric statistical
techniques is the use of the order (or 'ranks') of the data instead of
their actual values \citep[e.g.][]{Leh1998}. The use of ranks instead
of the actual values can sometimes lead to surprisingly powerful
results, in spite of the relative simplicity of the calculation and
the obviously reduced information. Our hope is that this approach will
be appropriate in cases where we are not sure of either the underlying
periodic function or the statistical distribution of the noise.

We suggest to approach the problem of period detection through the
notion of 'randomness tests'. When we do not phase-fold the data or
phase-fold it in a wrong period, the data points are expected to
behave randomly, whereas the correct phase-folding should reduce this
randomness. Quantifying randomness is an important problem in the field
of cryptography, which is concerned, among other things, with
deterministically producing sequences that should exhibit randomness
qualities. Thus, cryptography literature is rich in statistical tests,
parametric and non-parametric, to test for randomness
\citep[e.g.][]{Ruk2001}. Some of the tests we present here are also
used in cryptography for this end.

Astronomical surveys have different characteristics, with respect to
their noise distribution, sampling cadence, and size. In the current
paper we compare the performance of the various techniques we
introduce, when applying them to time series that contain a few dozen
points, with an uneven sampling law. Basically, the performance tests
we applied assume a survey with sampling characteristics that are very
roughly similar to \textit{Hipparcos} \citep{ESA1997} or
\textit{Gaia} \citep{Jor2008}, {only in terms of their size and mean
cadence}.  Since the data have to be phase-folded for every trial
period, time series longer than a few thousand points, like the light
curves of \textit{CoRoT} \citep{Auvetal2009} or
\textit{Kepler} \citep{Kocetal2010}, would probably require immense
computing power and might lose their practicality.

It is important to mention that there have already been several
attempts to apply the non-parametric approach to the problem of period
detection in astronomical time series. However, the previous works use
a somewhat different tool of non-parametric statistics --
non-parametric regression. Those works attempt not only to estimate
the period, but also the shape of the periodic function, under minimal
assumptions, e.g. using Gaussian Kernels. Such are the works by
\citet*{Haletal2000}, \citet{HalLi2006} and
\citet*{Wanetal2012}. Recently, \citet*{Sunetal2012} proposed another
non-parametric approach to period estimation, but it heavily relied on
even sampling. Those studies emphasize the performance of their
respective proposed methods in terms of estimated period accuracy. In
our case, since we focus on sparsely sampled time series, we do not
aim at the best achievable period accuracy, but at the ability to
merely detect the periodicity with as few data points as
possible. Such detection could trigger follow-up observations that
would augment the sparse data to allow a more accurate period
determination.

In the next section we introduce the non-parametric approaches that
are the topic of this work. Later, in Section \ref{simulations} we
detail the suite of simulations and tests we used to perform the
benchmark experiment. We discuss the results and offer some more
insights in the last section.

\section{Non-Parametric Measures of Serial Dependence}

After a scan of the literature about non-parametric statistics, we
chose five non-parametric methods to quantify the dependence between
two variates. In fact the literature contains much more methods but we
focused on five which we could tell were really independent.

Assuming there are $N$ data points, let us denote the phase-folded
data by $x_i$~($i=1,...,N$) where the index $i$ denotes the order
after the phase folding. A serial dependence measure will test the
dependence of the bivariate sample that consists of the pairs:
\begin{equation}
(x_1,x_2) , (x_2,x_3) , ... , (x_{N-1},x_N), (x_N,x_1)\ .
\end{equation}
Note the cyclic wraparound with the pair $(x_N,x_1)$. Some of the
non-parametric methods use the rank statistics, where each value is
replaced by its rank $R_i$, i.e., its place in the sorted sequence of
values. For example, the sequence:
\begin{equation} 
x_i = 2.3, 5.4, 3.2, 5.5, 5.6, -3.2, 0, 19.4, 1.2, 40
\end{equation}
will yield the rank sequence:
\begin{equation}
R_i = 4, 6, 5, 7, 8, 1, 2, 9, 3, 10
\end{equation}
Other methods go further in reducing the information by replacing the
value $x_i$ by a flag $F_i$ that signifies whether the actual value is
above or below the median of the sample. In the example above, the
flag sequence will read:
\begin{equation}
F_i = 0, 1, 0, 1, 1, 0, 0, 1, 0, 1
\end{equation}

The literature suggests various special ways to deal with cases of
ties or of odd sample size (where the median is actually one of the
values). Furthermore, it should be emphasized that all the appearances
in the literature of the tests we mention here are used originally as
general dependence tests, not meant specifically for period
detection. Thus, the original formulation of the tests uses no cyclic
wraparound.

\subsection{Bartels test}

Robert Bartels introduced this test in 1982 as a rank version of the
von Neumann's ratio test \citep{Bar1982}. In our case the expression
for computing Bartels' statistic is:
\begin{equation}
\beta = \sum_{i=1}^{N-1} (R_i-R_{i+1})^2 + (R_N-R_1)^2
\end{equation}
One may say that this statistic is essentially a string-length
statistic calculated for the ranks rather than for the actual values,
and indeed this is the way Bartels described it.

In the same way that von Neumann's ratio can be linearly related to
the serial correlation coefficient, Bartels' expression can be
linearly related to the serial Spearman's rank correlation coefficient
\citep{Spe1904}, and thus they are essentially equivalent. Note that
in this current formulation, similarly to the von-Neumann ratio
statistic, we expect a periodicity metric based on the Bartels
statistic to exhibit a trough rather than a peak at the correct
period.

\subsection{Kendall's tau ($\tau$)}

Maurice Kendall introduced this test in 1938 \citep{Ken1938}, as a
measure of rank correlation. The calculation of Kendall's $\tau$ is
rather cumbersome. In order to calculate $\tau$ one goes over all
$\frac{1}{2} N (N-1)$ pairs of ordered pairs of data points, i.e.,
pairs of the form $\left( (x_i,x_{i+1}) , (x_j,x_{j+1}) \right)$. Of
course, we assume that the ordered pairs include the pair $(x_N,x_1)$,
in order to treat correctly the wraparound.  A pair is said to be
concordant if the ranks for both elements agree, i.e., if both
$x_i>x_{i+1}$ and $x_j>x_{j+1}$, or if both $x_i<x_{i+1}$ and
$x_j<x_{j+1}$.  Otherwise it is said to be discordant. Denote by $N_c$
the number of concordant pairs and by $N_d$ the number of discordant
pairs. Kendall's $\tau$ is then defined by:
\begin{equation}
\tau = \frac{N_c-N_d}{\frac{1}{2} N (N-1)}\ .
\end{equation}

Kendall's $\tau$ is non-parametric in the sense that the actual values
of the data are not relevant, only their order.

\subsection{Runs test}

The runs test procedure counts the number of runs of constant flags
$F_i$, or equivalently, the number of times the flags change. Recall
that the flag $F_i$ equals zero if the original value is below the
median, and equals one otherwise. Thus, one may write an algebraic
definition:
\begin{equation}
U = \sum_{i=1}^{N-1} |F_{i+1}-F_i| + |F_1-F_N|
\end{equation}

\citet{WalWol1940} first introduced this test in a somewhat different
context of testing whether two samples were drawn from the same
population. In our case, obviously, runs are counted only for a single
cycle. Note that at the correct period we expect the data points to
get organized in as few runs as possible, so we expect the correct
period to exhibit a trough in the periodicity metric plot rather than
a peak (like in the Bartels-test or the von-Neumann-ratio cases).

\subsection{Corner test}

\citet{OlmTuk1947} proposed the rather elaborate corner test for
finding association between two variables, in situations where we
suspect that information about association may concentrate in points
on the periphery of the dataset. To effect the test, in the $xy$
scatter plot of the paired dataset, draw in the $x$ and $y$ median
lines, and label the quadrants so formed $+$,$-$,$+$,$-$, serially
from the top right-hand corner. Thus, in this new Cartesian coordinate
system, the first and third quadrants are labeled positive, and the
second and fourth negative.  Then, starting at the top, move
vertically down counting points with decreasing $y$ values until it is
necessary to cross the $x$ median line, and attach to this count the
sign of the quadrant in which the points lie. Proceed in similar
fashion for the bottom, left- and right-hand sides of the dataset. The
test statistic is then the sum of the four counts with their
signs. \cite{OlmTuk1947} provide an example for the calculation,
together with a graphic visualization, and suggest a treatment for
ties.

\subsection{Hoeffding test}

In 1948, Wassily Hoeffding proposed an even more elaborate statistic
than the corner test \citep{Hoe1948}. Hoeffding's statistic is based
on the population measure of the deviation of the bivariate
distribution from independence. This statistic is especially suitable
for quantifying general kinds of dependence, not necessarily linear or
even monotone. Adapted to our case including the wraparound points,
Hoeffding's statistic is computed as follows:
\begin{equation}
\label{HoeffD}
D = \frac{A-2(N-2)B+(N-2)(N-3)C}{N(N-1)(N-2)(N-3)(N-4)}
\end{equation}
where
\begin{dmath}
\label{HoeffA}
A = \sum_{i=1}^{N-1} (R_i-1)(R_i-2)(R_{i+1}-1)(R_{i+1}-2) + (R_N-1)(R_N-2)(R_1-1)(R_1-2)\ ,
\end{dmath}
\begin{equation}
\label{HoeffB}
B = \sum_{i=1}^{N-1} (R_i-2)(R_{i+1}-2)c_i +(R_N-2)(R_1-2)c_N\ ,
\end{equation}
\begin{equation}
\label{HoeffC}
C = \sum_{i=1}^N c_i(c_i-1) .
\end{equation}
where $c_i$ (sometimes called the bivariate rank) is the number of
pairs $(x_j,x_{j+1})$ for which both $x_j<x_i$ and
$x_{j+1}<x_{i+1}$. 

\section{Simulations}
\label{simulations}

We present here a suite of simulations we have performed in order to
assess the competitiveness of the tests introduced in the previous
section against the traditional methods of least squares and string
length.

\subsection{Simulated signals}

In all the simulations we present here, we randomly drew a sparse set
of sampling times, from a total time baseline spanning $1000$ time
units (for convenience, we henceforth denote our arbitrary time unit
by 'days', referring to the intended applications in astronomy). We
then used those times to sample a periodic function (see below) and
added white Gaussian noise, whose amplitude we determined based on a
prescribed signal to noise ratio (SNR). The SNR here means the ratio
between the standard deviation of the specific simulated data points
without the noise, and the standard deviation of the Gaussian
noise. In the simulations we present here we simulated a periodic
function with a period of two days. We checked with a few simulations
that the results did not change in any significant manner when we
tried other periods.

The simulations are intended to be very generic. We wished to
test the case of an unevenly sampled sparse dataset, i.e. with not
many samples. We therefore used the most generic non-uniform sampling
-- random times drawn from a uniform distribution. We thus have not
included any sampling biases one might encounter in astronomy. We also
used the most generic form of noise -- noise drawn from a Gaussian
distribution. When the techniques we present here are implemented in
specific astronomical cases, it will be important to test them in the
specific context.

The different shapes of periodic functions we used, ranged from purely
sinusoidal shape to extremely non-sinusoidal. The shapes and the
formulae we used to produce them are detailed below, as a function of
the phase ${\phi=\frac{t\mod{P}}{P}}$\ :

\begin{description}

\item[A.] Pure sinusoidal function:
\begin{equation}
x = \cos(2\pi\phi)
\end{equation}

\item[B.] Mildly non-sinusoidal function ('almost sinusoidal')	:
\begin{equation}
x = \cos(2\pi\phi) + 0.25\cos(4\pi\phi)
\end{equation}

\item[C.] Sawtooth wave:
\begin{equation}
x = \phi
\end{equation}

\item[D.] Pulse wave. In the simulations we have used a pulse wave with
pulse duration of two thirds of the period:
\begin{equation}
x = \begin{cases}
	0& \text{if $\phi<1/3$},\\
	1& \text{otherwise}.
\end{cases}
\end{equation}

\item[E.] Eclipsing binary light curve. We approximated the lightcurve
of an eccentric eclipsing binary by subtracting two Gaussian functions
from a constant baseline of $1$. The first Gaussian was centred around
phase $0.0$, with a maximum value of $0.5$, and the second one centred
around phase $0.4$, with a maximum value of $0.2$.

\item[F.] Spectroscopic binary radial velocity (RV) curve. Using the
well-known formulae for spectroscopic binary RV, we simulated a RV
curve with an eccentricity of $0.97$ and an argument of periastron
$\omega=\frac{\pi}{4}$.

\end{description}

\begin{figure}
\includegraphics[width=0.5\textwidth]{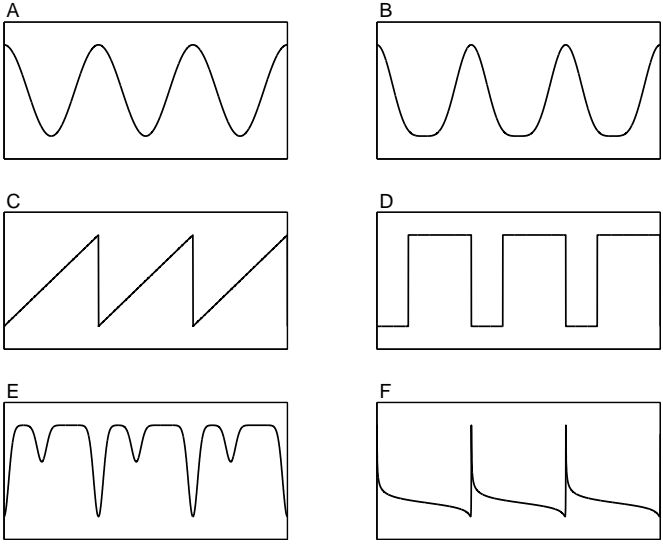}
\caption{A Schematic illustration of the six periodic functions used in
the simulations. A.\ Sinusoidal; B.\ Almost sinusoidal; C.\ Sawtooth;
D.\ Pulse wave; E.\ Eclipsing binary light curve; F.\ Eccentric
spectroscopic binary RV curve. For detailed descriptions of the
various shapes see text.}
\label{shapes}
\end{figure}

The exact amplitudes and offset values of the above detailed functions
do not matter for our purposes, only the SNR of the various cases we
simulated. Fig.\ \ref{shapes} portrays schematically the six shapes.

\subsection{Examples}

\begin{figure}
\includegraphics[width=0.5\textwidth]{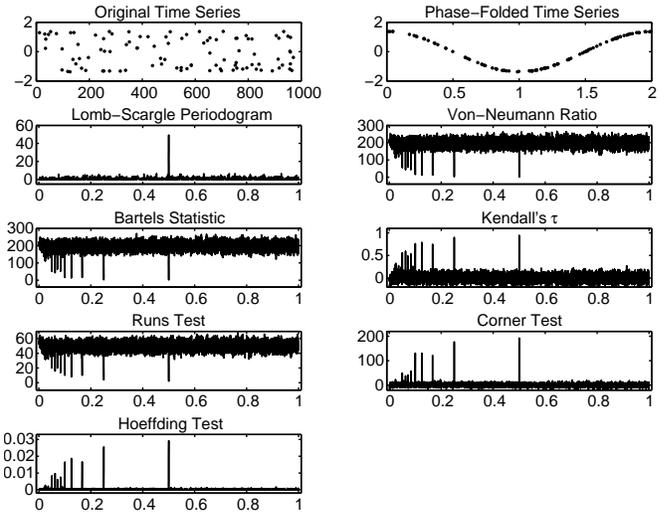}
\caption{The results of applying all the examined periodicity
detection methods to a simulated time-series, of the sinusoidal type,
with $100$ simulated data points, and a SNR of $100$.  The upper two
panels show the time series and its phase-folded version, using the
correct period. The other panels show the periodicity metrics
calculated for this time series, with self explanatory titles.}

\label{SN_100}
\end{figure}

\begin{figure}
\includegraphics[width=0.5\textwidth]{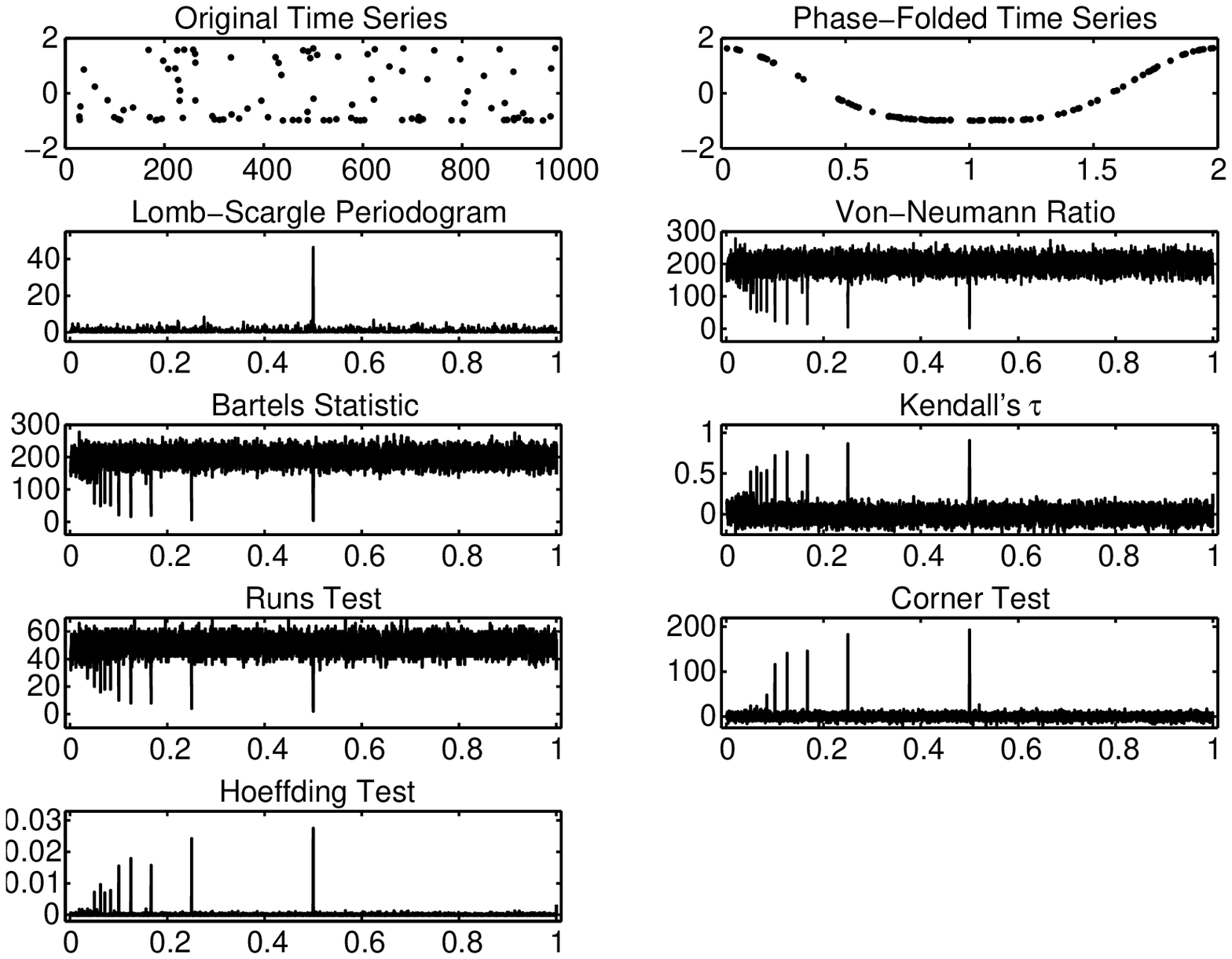}
\caption{The results of applying all the examined periodicity
detection methods to a simulated time-series, of the 'almost
sinusoidal' type, with $100$ simulated data points, and a SNR of
$100$.  The upper two panels show the time series and its phase-folded
version, using the correct period. The other panels show the
periodicity metrics calculated for this time series, with self
explanatory titles.}
\label{AS_100}
\end{figure}

\begin{figure}
\includegraphics[width=0.5\textwidth]{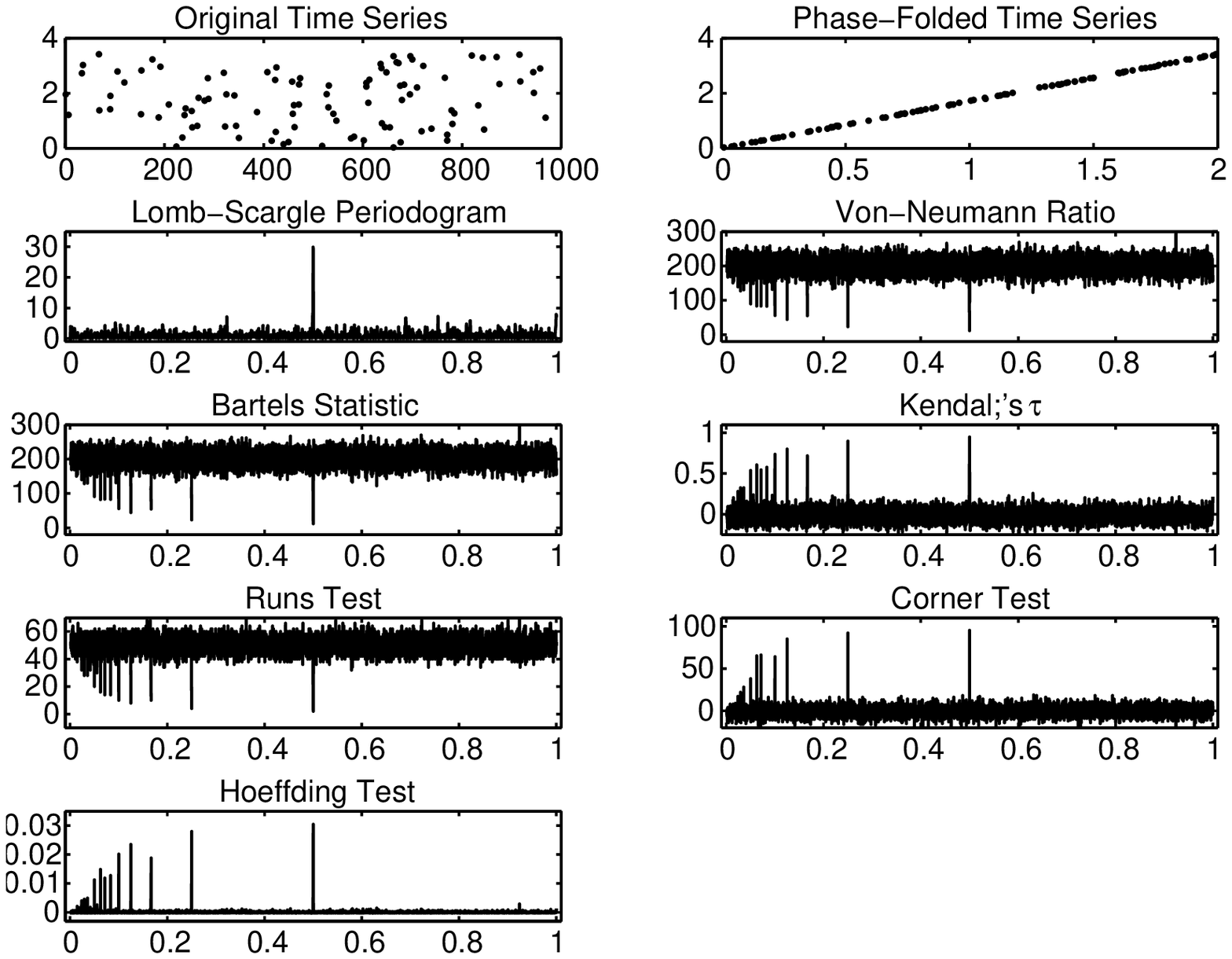}
\caption{The results of applying all the examined periodicity
detection methods to a simulated time-series, of the sawtooth type,
with $100$ simulated data points, and a SNR of $100$.  The upper two
panels show the time series and its phase-folded version, using the
correct period. The other panels show the periodicity metrics
calculated for this time series, with self explanatory titles.}
\label{ST_100}
\end{figure}

\begin{figure}
\includegraphics[width=0.5\textwidth]{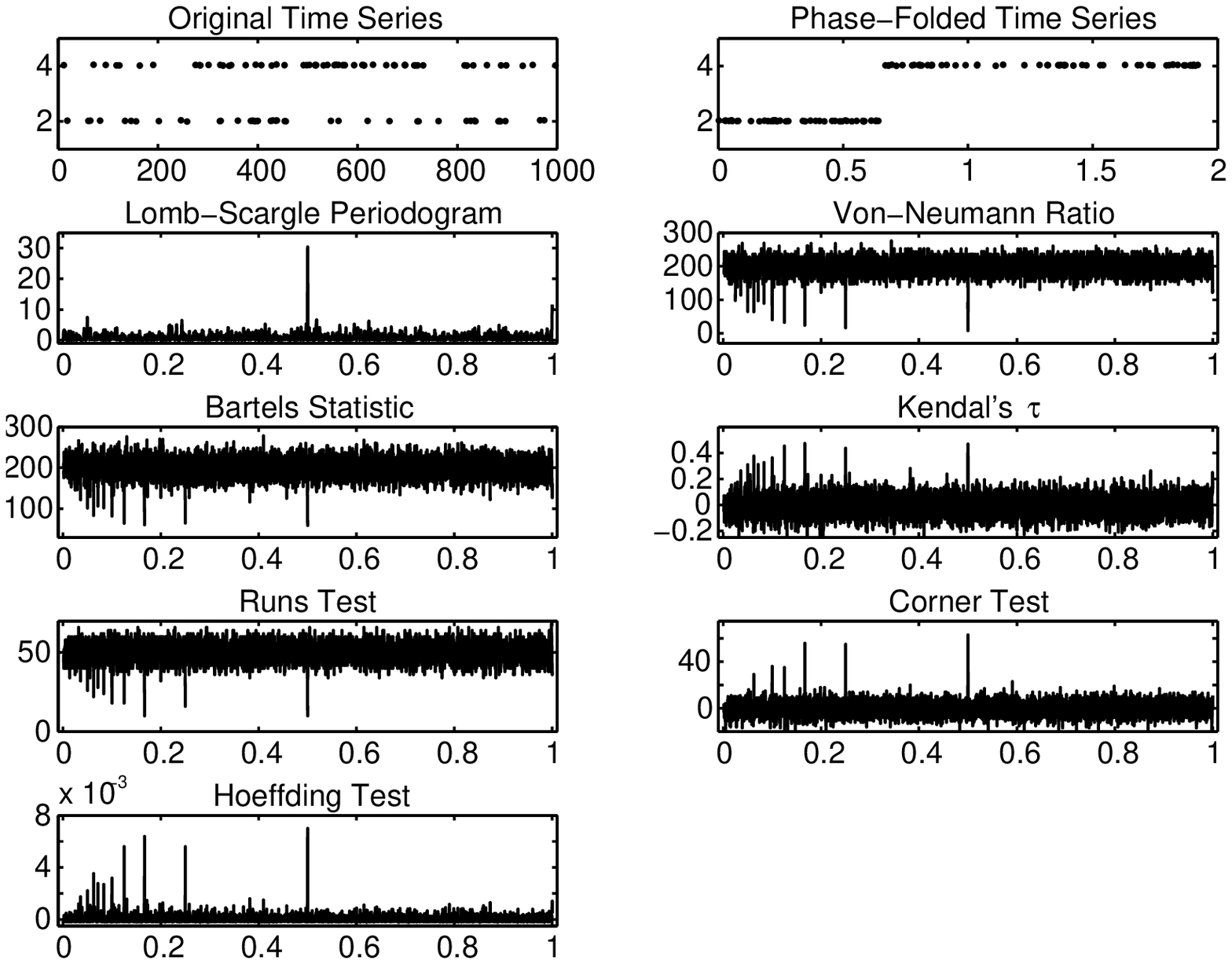}
\caption{The results of applying all the examined periodicity
detection methods to a simulated time-series, of the pulse wave type,
with $100$ simulated data points, and a SNR of $100$.  The upper two
panels show the time series and its phase-folded version, using the
correct period. The other panels show the periodicity metrics
calculated for this time series, with self explanatory titles.}
\label{PL_100}
\end{figure}

\begin{figure}
\includegraphics[width=0.5\textwidth]{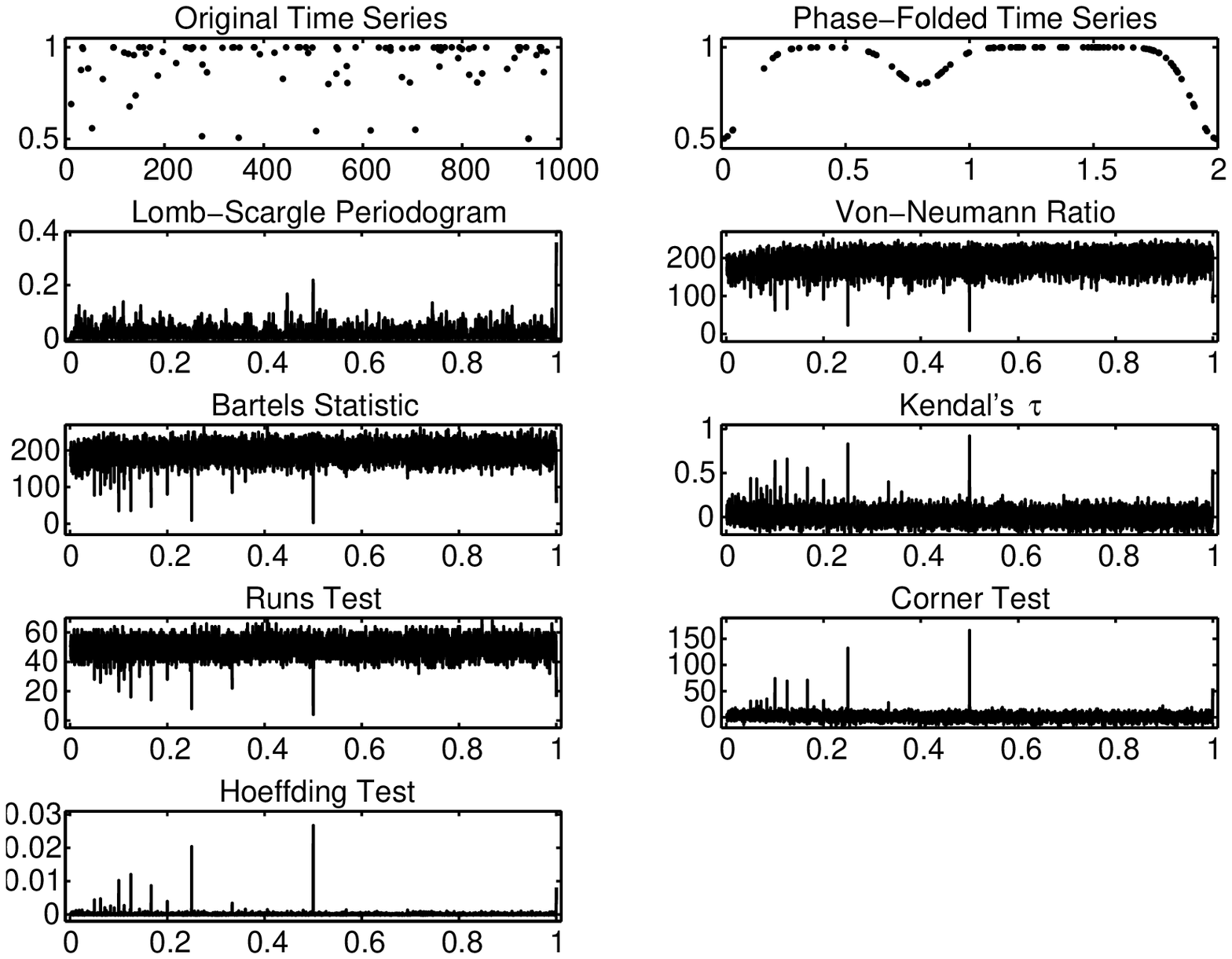}
\caption{The results of applying all the examined periodicity
detection methods to a simulated time-series, of the eclipsing binary
lightcurve type, with $100$ simulated data points, and a SNR of $100$.
The upper two panels show the time series and its phase-folded
version, using the correct period. The other panels show the
periodicity metrics calculated for this time series, with self
explanatory titles.}
\label{EB_100}
\end{figure}

\begin{figure}
\includegraphics[width=0.5\textwidth]{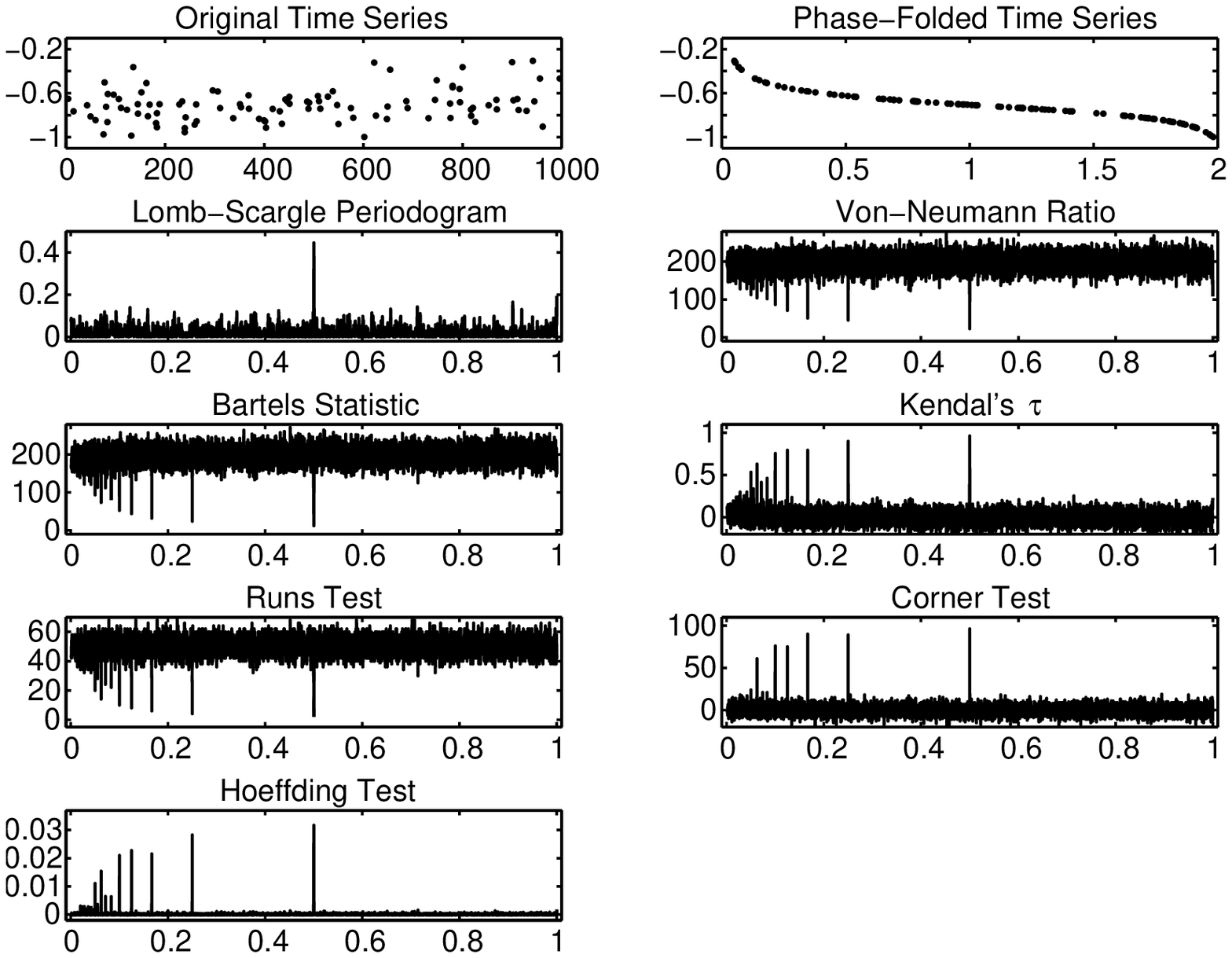}
\caption{The results of applying all the examined periodicity
detection methods to a simulated time-series, of the spectroscopic
binary RV curve type, with $100$ simulated data points, and a SNR of
$100$.  The upper two panels show the time series and its phase-folded
version, using the correct period. The other panels show the
periodicity metrics calculated for this time series, with self
explanatory titles.}
\label{SB_100}
\end{figure}

Figs \ref{SN_100}--\ref{SB_100} show the results of applying all the
methods on simulated periodic time series with $100$ data points, and
SNR of $100$. This is a relatively easy task, and in almost all cases,
the correct peak is retrieved, except for the case of the
Lomb--Scargle periodogram applied to the eclipsing binary case. In
this specific case, the Lomb--Scargle periodogram has a prominent peak
in the second harmonic of the true frequency (at a frequency of
$1\,\mathrm{d}^{-1}$), instead of the correct one. The peak in the
correct period is much lower than this second harmonic. This is a well
known feature of the Lomb--Scargle periodogram, and it usually does
not constitute a problem, since the peak at the second harmonic
already leads to a detection, and further scrutiny of the light curve
reveals the correct period. One can also clearly see the appearance of
sub-harmonic peaks at almost all periodicity metric plots, besides
Lomb--Scargle periodogram. Judging subjectively only by these specific
idealized examples, we can say that besides the case of the pulse-wave
periodicity, the Hoeffding-test periodicity metric provides the
'cleanest' detection: a clear peak at the correct period, with very
low variability in other periods (potentially implying a very small
chance of false detection), except for the subharmonics.

\begin{figure}
\includegraphics[width=0.5\textwidth]{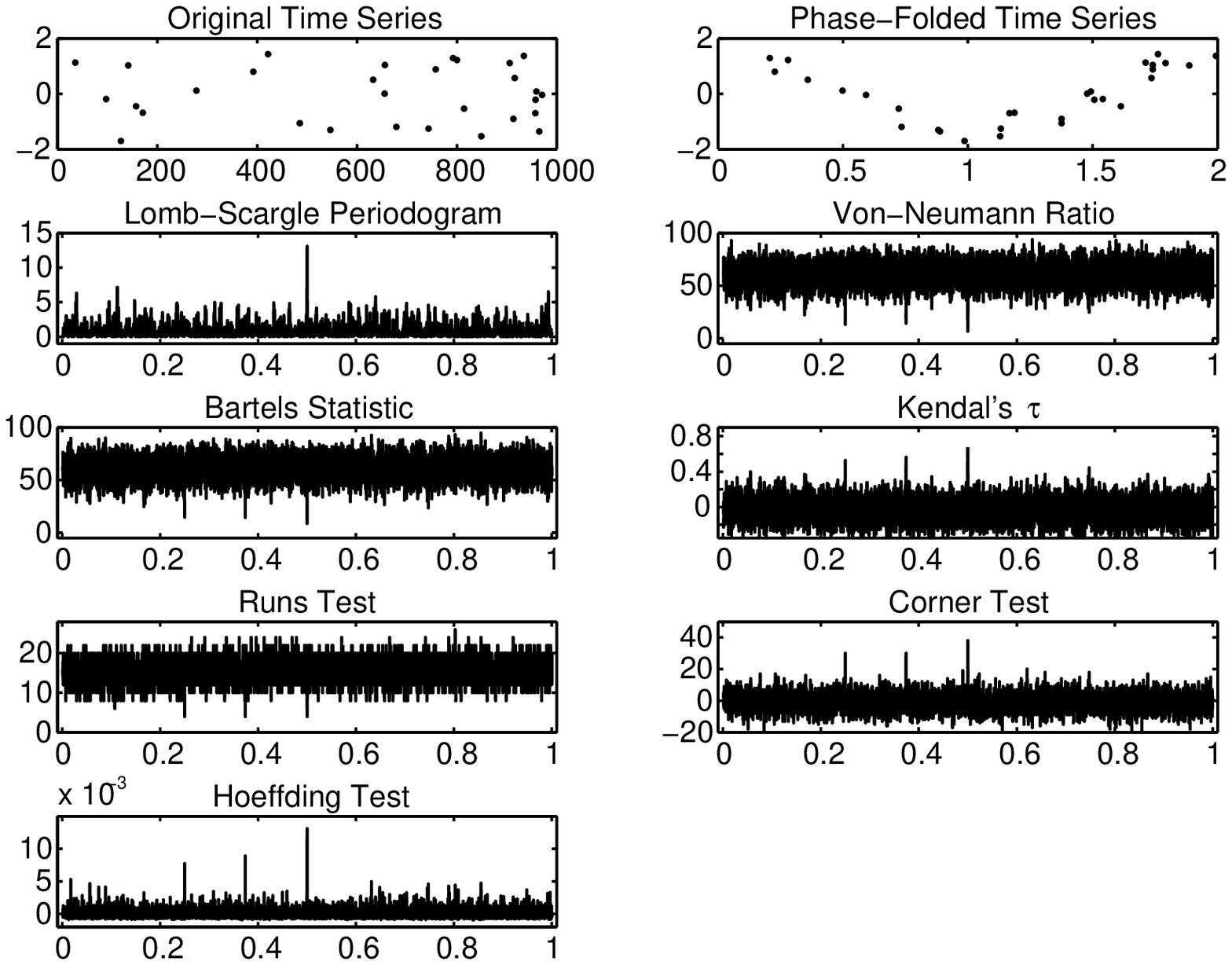}
\caption{The results of applying all the examined periodicity
detection methods to a simulated time-series, of the sinusoidal type,
with $30$ simulated data points, and a SNR of $3$.  The upper two
panels show the time series and its phase-folded version, using the
correct period. The other panels show the periodicity metrics
calculated for this time series, with self explanatory titles.}
\label{SN_30}
\end{figure}

\begin{figure}
\includegraphics[width=0.5\textwidth]{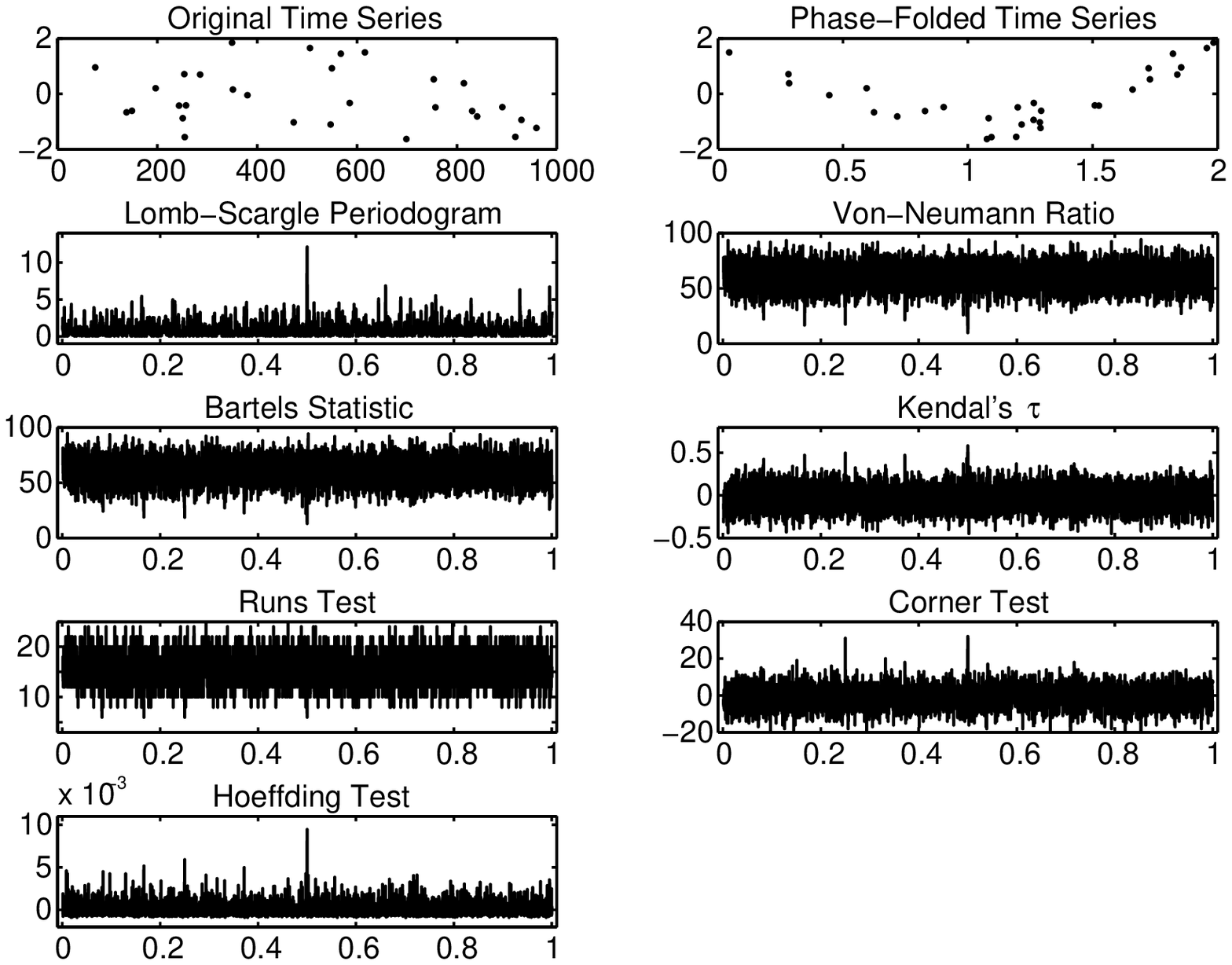}
\caption{The results of applying all the examined periodicity
detection methods to a simulated time-series, of the 'almost
sinusoidal' type, with $30$ simulated data points, and a SNR of $3$.
The upper two panels show the time series and its phase-folded
version, using the correct period. The other panels show the
periodicity metrics calculated for this time series, with self
explanatory titles.}
\label{AS_30}
\end{figure}

\begin{figure}
\includegraphics[width=0.5\textwidth]{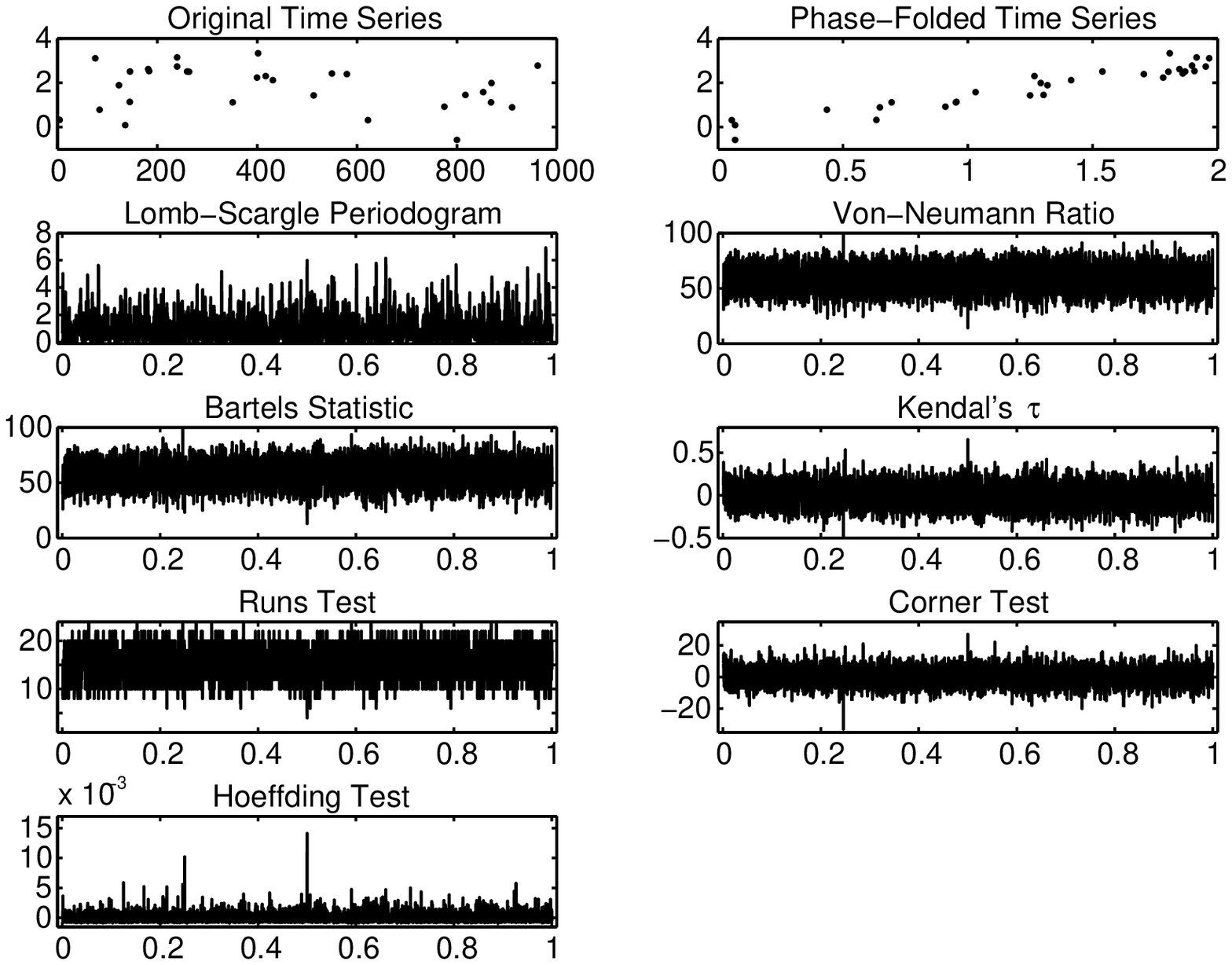}
\caption{The results of applying all the examined periodicity
detection methods to a simulated time-series, of the sawtooth type,
with $30$ simulated data points, and a SNR of $3$.  The upper two
panels show the time series and its phase-folded version, using the
correct period. The other panels show the periodicity metrics
calculated for this time series, with self explanatory titles.}
\label{ST_30}
\end{figure}

\begin{figure}
\includegraphics[width=0.5\textwidth]{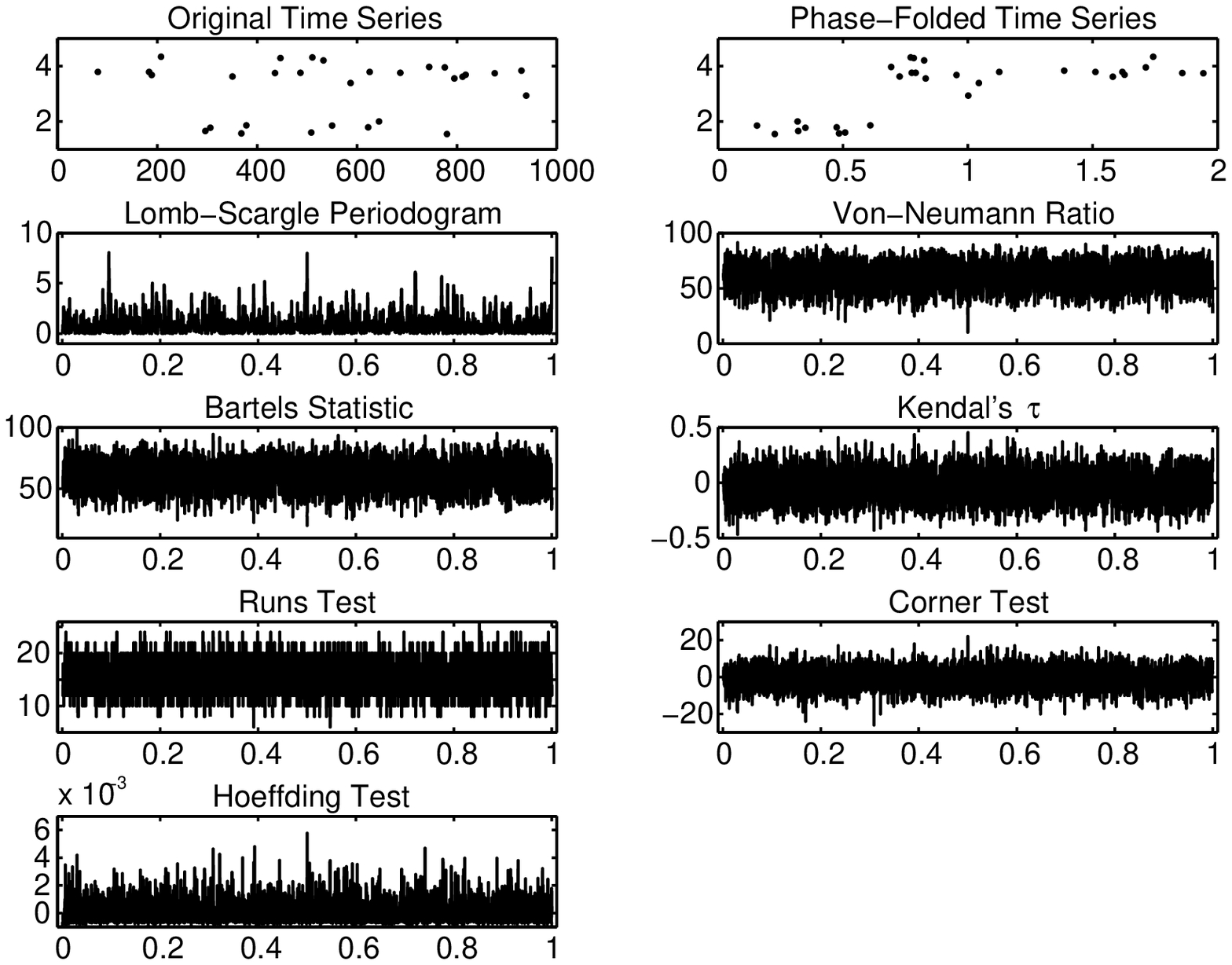}
\caption{The results of applying all the examined periodicity
detection methods to a simulated time-series, of the pulse wave type,
with $30$ simulated data points, and a SNR of $3$.  The upper two
panels show the time series and its phase-folded version, using the
correct period. The other panels show the periodicity metrics
calculated for this time series, with self explanatory titles.}
\label{PL_30}
\end{figure}

\begin{figure}
\includegraphics[width=0.5\textwidth]{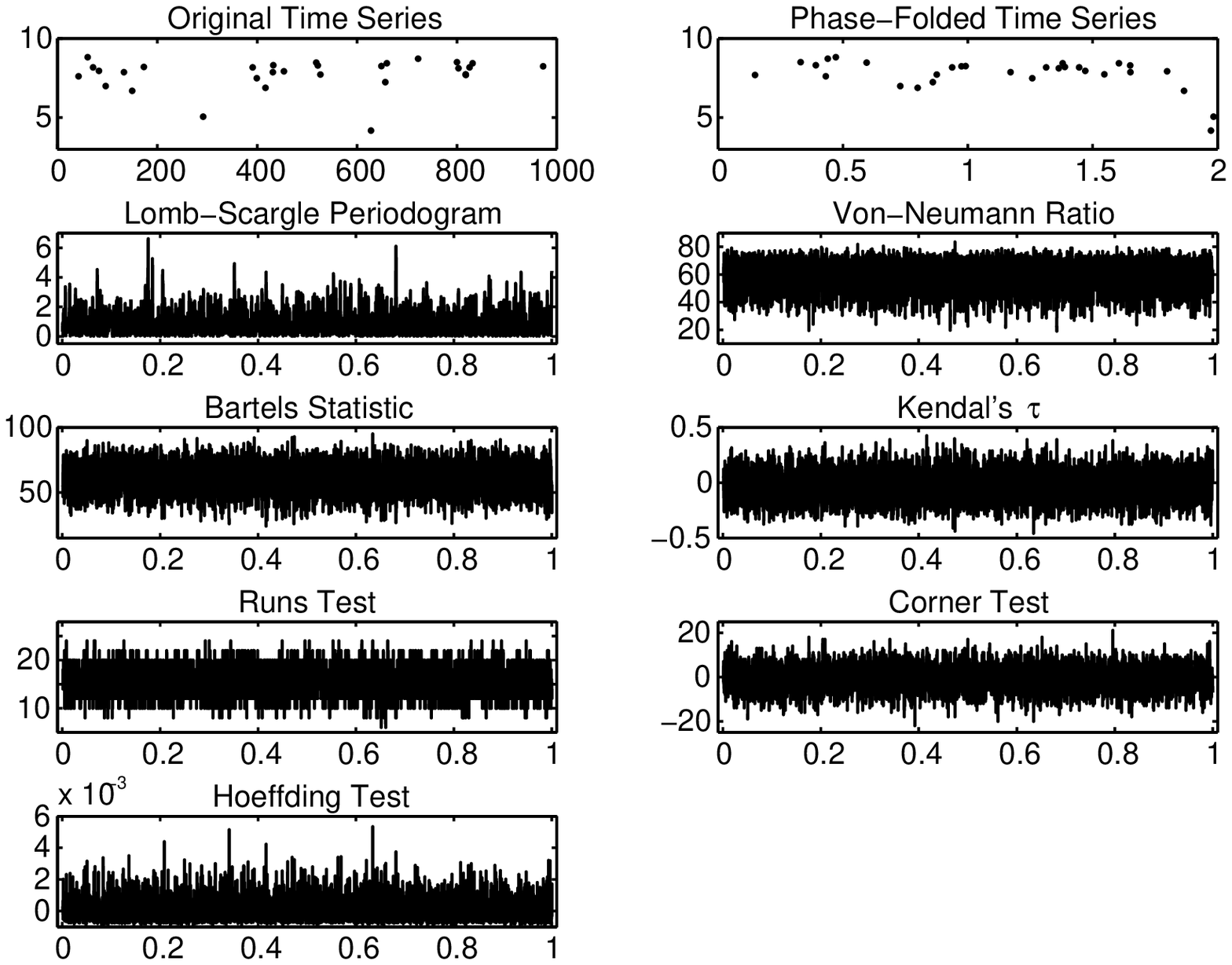}
\caption{The results of applying all the examined periodicity
detection methods to a simulated time-series, of the eclipsing binary
light curve type, with $30$ simulated data points, and a SNR of $3$.
The upper two panels show the time series and its phase-folded
version, using the correct period. The other panels show the
periodicity metrics calculated for this time series, with self
explanatory titles.}
\label{EB_30}
\end{figure}

\begin{figure}
\includegraphics[width=0.5\textwidth]{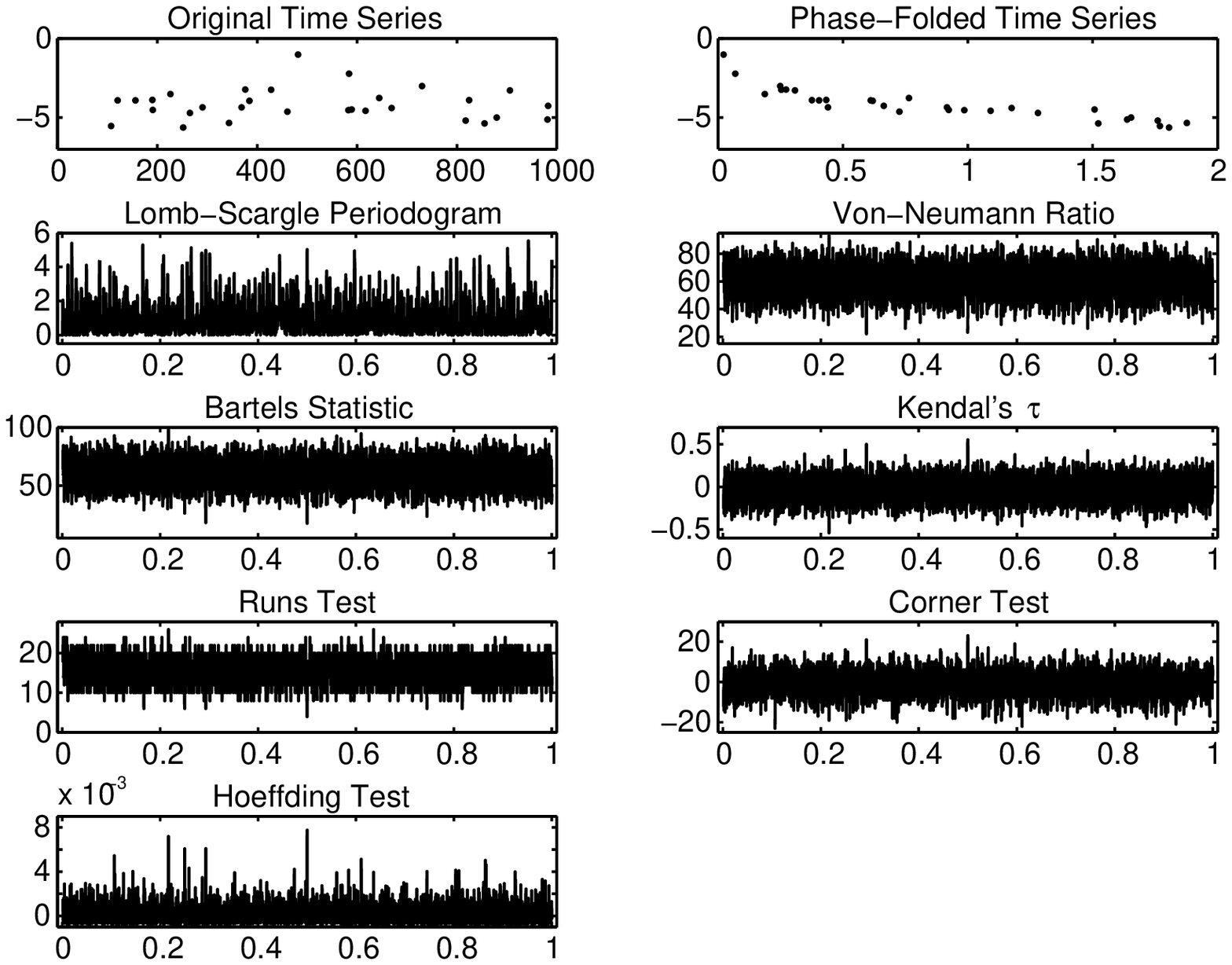}
\caption{The results of applying all the examined periodicity
detection methods to a simulated time-series, of the spectroscopic
binary RV curve type, with $30$ simulated data points, and a SNR of
$3$.  The upper two panels show the time series and its phase-folded
version, using the correct period. The other panels show the
periodicity metrics calculated for this time series, with self
explanatory titles.}
\label{SB_30}
\end{figure}

Figs \ref{SN_30}--\ref{SB_30} present tougher cases -- time series
with $30$ data points, and SNR of $3$. This time, as one would expect,
the results are not as impressive as in the cases with more points and
higher SNR. In the sinusoidal and almost-sinusoidal cases it seems
that all metrics still exhibit the correct peak, but it is much less
prominent against the background of the other periods. The
Hoeffding-test metric still competes very well with the Lomb--Scargle
periodogram, and seems to be much more conclusive compared to the
other methods. In the sawtooth and the spectroscopic binary case the
superiority of the Hoeffding-test metric over all the others,
including the Lomb--Scargle periodogram, is evident. In the pulse-wave
and the eclipsing binary cases all methods perform poorly, with a
minor preference for Lomb--Scargle over the Hoeffding test.

\subsection{Performance testing}

For each simulated time series we calculated the five periodicity
metric functions we introduced earlier. As a base for comparison we
also calculated the classical Lomb--Scargle periodogram, and also the
von-Neumann ratio {periodicity metric}, as representatives of the more
traditional methods.

We calculated the {periodicity metric functions} for a grid of
frequencies, ranging from $10^{-4}$ to $1\,\mathrm{d}^{-1}$, with
steps of $10^{-4}\,\mathrm{d}^{-1}$. Thus, the simulated period of two
days is located at the middle of the frequency range we used, at a
frequency of $0.5\,\mathrm{d}^{-1}$.

Knowing the true period, we used as a performance metric for our
benchmark experiment, the degree to which the periodicity score
statistic at the correct period can be singled out compared to the
values at other periods. This should be made cautiously, since the
null-hypothesis distribution is different for each statistic. In order
to overcome this hurdle we prepared for each statistic a reference
sample of random datasets (pure noise, no signal) of the same number
of points as the one being tested. The size of the reference sample
was $10^6$. Each value in the tested method was converted using
quantiles of this empirical reference distribution. In order to
transform extreme values, that were not obtained in the random $10^6$
reference samples, we extrapolated the distribution using a
maximum-likelihood fit of a generalized Pareto distribution that was
applied to the $10^{-3}$ distribution tail
\citep[e.g.][]{deZKot2010}. Our experience showed that robustness 
required to exclude the $10$ most extreme values of the sample, while
constraining the Pareto distribution shape parameter to be
non-positive \citep{deZKot2010}. Generalized Pareto distribution was
not suitable for the runs-test distribution tail, but a Gaussian
approximation seemed to be a good-enough approximation (in the same
way it is used as an approximation to the binomial distribution).

After the conversion based on the reference distribution, we measured
the difference between the value at the correct period and the average
value of the statistic , and normalized it by the standard deviation
of the statistic on all periods. Following \citet*{Kovetal2002} and
\citet{Alcetal2000}, we called this quantity SDE (Signal Detection
Efficiency):
\begin{equation}
\mathrm{SDE} = \frac{S(f_0)-\bar{S}}{\mathrm{SD}(S)}
\end{equation}
where $S$ is the tested periodicity metric function, $f_0$ is the true
frequency, and $\bar{S}$ and $\mathrm{SD}(S)$ are the mean and
standard deviation of the function. Note that our SDE is a little
different from the SDE in the papers by \citet{Kovetal2002} and
\citet{Alcetal2000}, where they use the peak value of the function,
and not the value at the known period. Our version of the SDE tests
specifically the ability of the tested method to single out the
correct period, which is known in advance in a simulation context.

\subsection{Performance trends}

\begin{figure}
\includegraphics[width=0.5\textwidth]{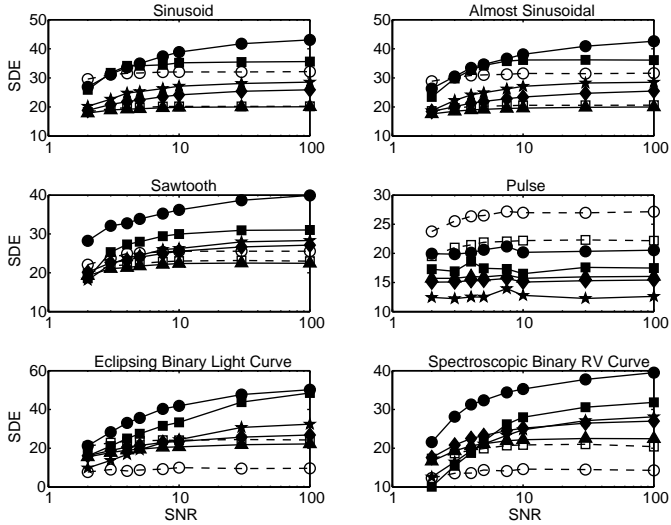}
\caption{SDE as a function of SNR for $100$ points time series. The
SDE is averaged over $100$ simulated light curves. Legend: empty
circles, dashed line -- Lomb--Scargle periodogram; empty squares,
dashed line -- von-Neumann Ratio; filled upward-pointing triangles,
solid line -- Bartels test; filled diamonds, solid line -- Kendall's
tau; filled pentagrams, solid line -- runs test; filled squares, solid
line -- corner test; filled circles, solid line -- Hoeffding-test .}
\label{N_100}
\end{figure}

In Fig.\ \ref{N_100} we present the results of performing $100$ time
series simulations with $100$ points each, and varying SNRs, for each
of the periodic functions. The values we plot in the figure are the
SDE, computed as we described above, after the conversion using the
reference distribution, averaged over the $100$ trials in every
configuration. In almost all situations the Hoeffding-test method
outperforms all other methods tested. The only situation where
it seems that the Lomb--Scargle consistently outperforms the
Hoeffding-test method is the case of the pulse-wave periodic
function. This was evident already in the examples (Fig.\
\ref{PL_100}). Probably the piecewise constant nature of the signal
makes ranking information irrelevant, whereas ranking has no effect on
the least-square fit. This also reflects in the fact that the
performance of the non-parametric techniques hardly improves with
increasing SNR for the pulse-wave case.

Based on the examples (Fig.\ \ref{EB_100}), we can also understand how
Lomb--Scargle performed so poorly in the eclipsing-binary cases. The
correct period (the fundamental frequency) almost does not exhibit a
peak in the periodogram, only the second harmonic does. The serial
dependence of the phase-folded light curve still remains, which allows
the non-parametric methods to perform the way they do in the other
cases.

\begin{figure}
\includegraphics[width=0.5\textwidth]{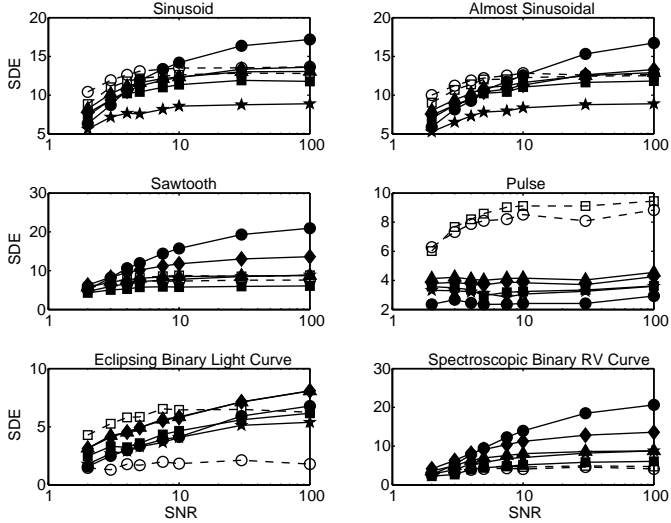}
\caption{SDE as a function of SNR for $20$ points time series. The
SDE is averaged over $100$ simulated light curves. Legend: empty
circles, dashed line -- Lomb--Scargle periodogram; empty squares,
dashed line -- von-Neumann Ratio; filled upward-pointing triangles,
solid line -- Bartels test; filled diamonds, solid line -- Kendall's
tau; filled pentagrams, solid line -- runs test; filled squares, solid
line -- corner test; filled circles, solid line -- Hoeffding-test.}
\label{N_20}
\end{figure}

In Fig.\ \ref{N_20} we present a more difficult case, where the number
of data points is much smaller -- $20$. This time the advantage of the
traditional techniques in the pulse-wave case is even more
pronounced. Some non-parametric techniques seem to be preferable over
the classical ones in the higher SNRs, and the Hoeffding test
maintains its superiority in most cases.

\begin{figure}
\includegraphics[width=0.5\textwidth]{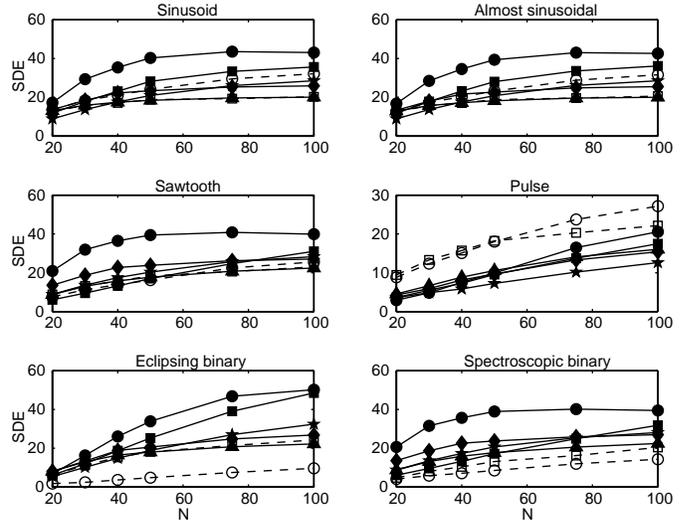}
\caption{SDE as a function of the number of data points for time
series with SNR of $100$. The SDE is averaged over $100$ simulated
light curves. Legend: empty circles, dashed line -- Lomb--Scargle
periodogram; empty squares, dashed line -- von-Neumann Ratio; filled
upward-pointing triangles, solid line -- Bartels test; filled
diamonds, solid line -- Kendall's tau; filled pentagrams, solid line
-- runs test; filled squares, solid line -- corner test; filled
circles, solid line -- Hoeffding-test.}
\label{SNR_100}
\end{figure}

Fig.\ \ref{SNR_100} examines the dependence of the SDE on the number
of points in the time series. This time we held the SNR fixed at $100$
while we varied the number of data points. Again, the figure displays
the averaged SDE over $100$ random realizations of the simulation. The
superiority of the Hoeffding-test method is evident for all kinds of
periodicities except for the pulse-wave, where the traditional methods
keep on being superior .

\begin{figure}
\includegraphics[width=0.5\textwidth]{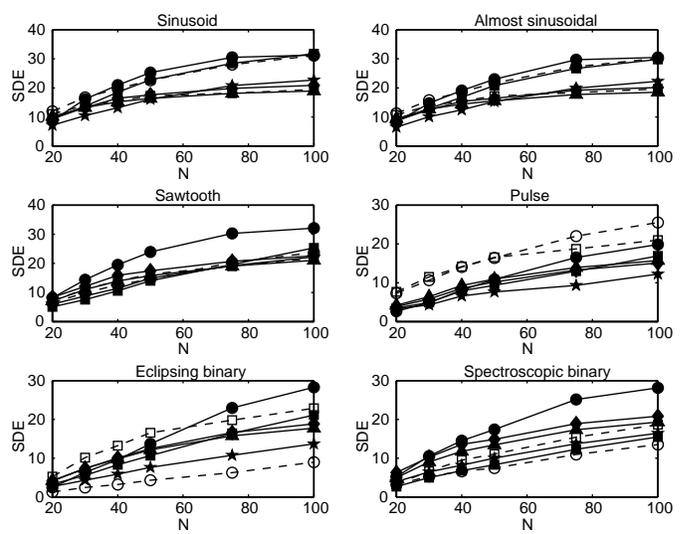}
\caption{SDE as a function of the number of data points for time
series with SNR of $3$. The SDE is averaged over $100$ simulated
light curves. Legend: empty circles, dashed line -- Lomb--Scargle
periodogram; empty squares, dashed line -- von-Neumann Ratio; filled
upward-pointing triangles, solid line -- Bartels test; filled
diamonds, solid line -- Kendall's tau; filled pentagrams, solid line
-- runs test; filled squares, solid line -- corner test; filled
circles, solid line -- Hoeffding-test.}
\label{SNR_3}
\end{figure}

Fig.\ \ref{SNR_3} presents the same test for the case of SNR fixed at
$3$. Again, the Hoeffding-test method is definitely superior for
the sawtooth wave and the spectroscopic binary RV case, and it is
competitive with the classical approaches in the other cases, except
the pulse wave.

We can summarize that unless there are significantly long constant
intervals in the periodicity shape, the superiority of the
Hoeffding-test method is most prominent in the extremely
non-sinusoidal cases, which are obviously of interest in astronomy --
the cases of the eclipsing binary light curve and the spectroscopic
binary RV curve.

\begin{figure}
\includegraphics[width=0.5\textwidth]{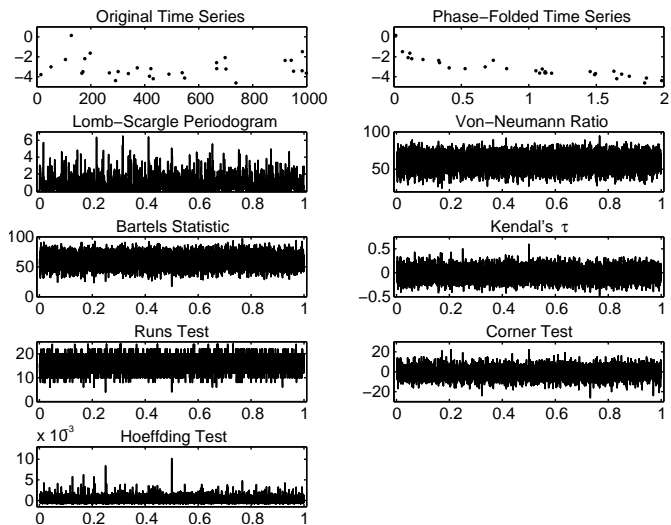}
\caption{An example of the results of applying all the examined
periodicity detection methods to a simulated time-series, of the
spectroscopic binary RV curve type, with $30$ simulated data
points, and a SNR of $3$.  The upper two panels show the time series
and its phase-folded version, using the correct period. The other
panels show the periodicity metrics calculated for this time series,
with self explanatory titles. Note the complete non-detection at the
Lomb--Scargle periodogram and the von-Neumann ratio, and
the very noticeable detection by the Hoeffding-test technique.}
\label{EX1}
\end{figure}

\begin{figure}
\includegraphics[width=0.5\textwidth]{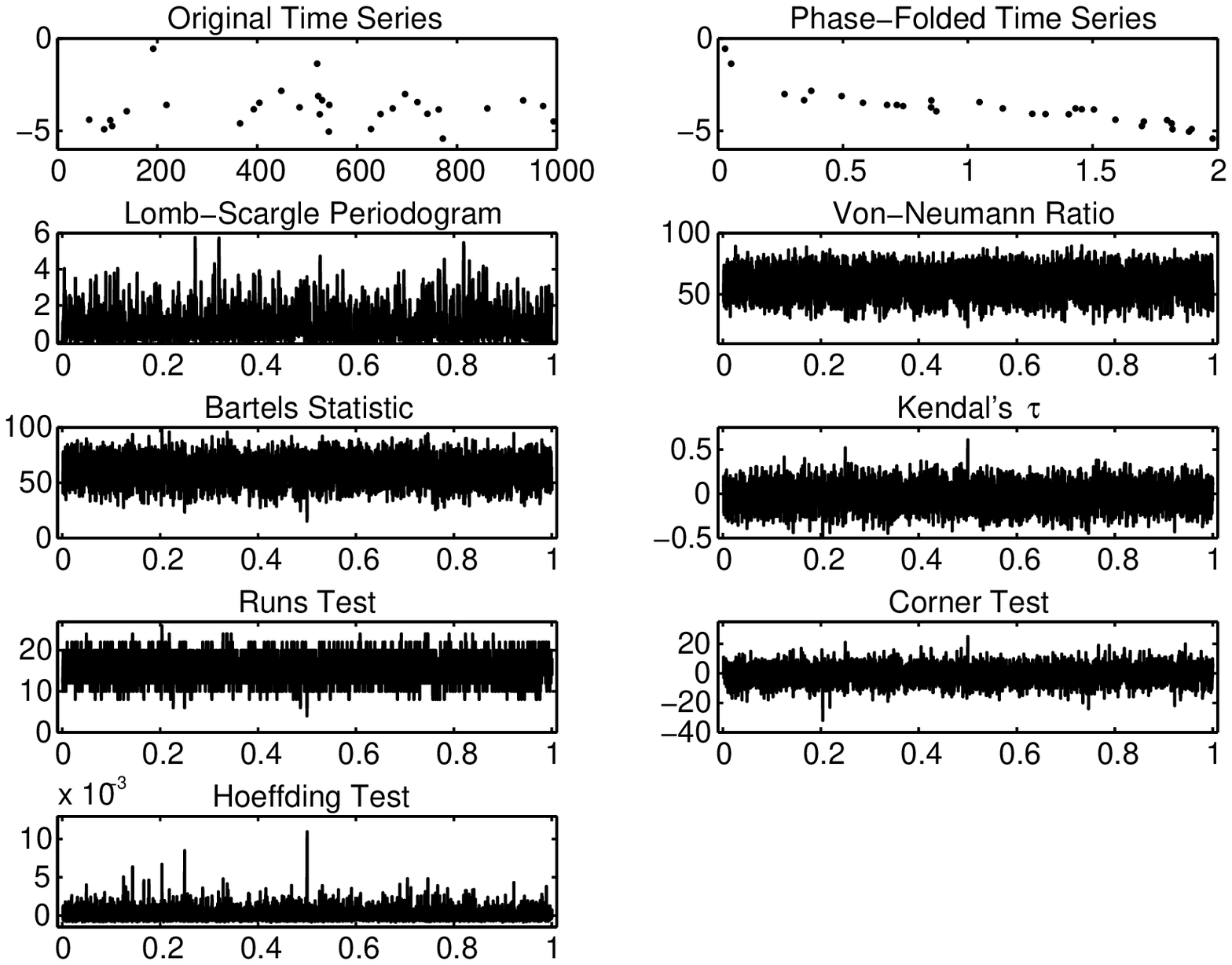}
\caption{Another example of the results of applying all the examined
periodicity detection methods to a simulated time-series, of the
spectroscopic binary RV curve type, with $30$ simulated data
points, and a SNR of $3$.  The upper two panels show the time series
and its phase-folded version, using the correct period. The other
panels show the periodicity metrics calculated for this time series,
with self explanatory titles. Note the complete non-detection at the
Lomb--Scargle periodogram and the von-Neumann ratio, and
the very noticeable detection by the Hoeffding-test technique.}
\label{EX2}
\end{figure}

\begin{figure}
\includegraphics[width=0.5\textwidth]{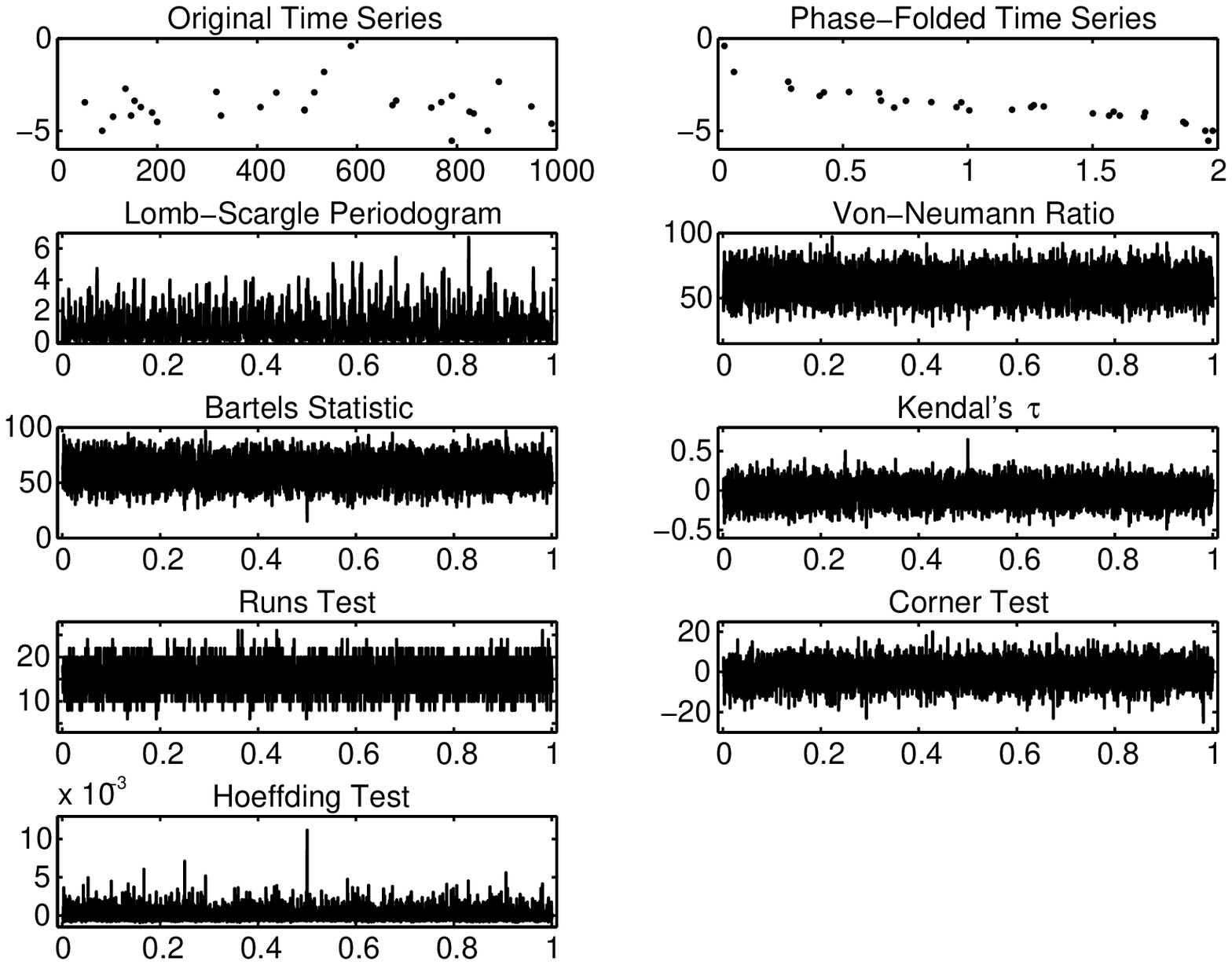}
\caption{Another example of the results of applying all the examined
periodicity detection methods to a simulated time-series, of the
spectroscopic binary RV curve type, with $30$ simulated data
points, and a SNR of $3$.  The upper two panels show the time series
and its phase-folded version, using the correct period. The other
panels show the periodicity metrics calculated for this time series,
with self explanatory titles. Note the complete non-detection at the
Lomb--Scargle periodogram and the von-Neumann ratio, and the very
noticeable detection by the Hoeffding-test technique.}
\label{EX3}
\end{figure}

The SDE plots we presented above are of a statistical nature. One may
still wonder whether those results have practical meaning.  We
therefore checked individually the spectroscopic binary RV simulations
for the case of $30$ data points and SNR of $3$ to get an
impression. According to Fig.\ \ref{SNR_3}, in this setting it seems
there was a marked difference between the performance of the Hoeffding
test and the classical methods. Indeed, it was quite easy to find
cases where the Lomb--Scargle periodogram simply missed the detection
while the Hoeffding-test method provided a clear detection (in fact,
they were the majority). We present three such example in Figs
\ref{EX1}--\ref{EX3}. This is obviously not a statistical proof, just
a demonstration of one aspect of the statistical study presented
earlier.

\section{Discussion}

We have presented here several kinds of periodicity metric statistics
based on non-parametric serial dependence measures, and assessed their
performance. It turns out that one of those methods, the
Hoeffding-test method, has the potential to be much more
powerful than the traditional parametric approaches of Lomb--Scargle
and string length for extremely non-sinusoidal periodicities.

The large time-domain surveys mentioned in the Introduction are
bound to provide, besides the easily detectable periodicity shapes,
also extreme cases with high harmonic content (non-sinusoidal). These
may be either extreme cases of eclipsing binaries, or even periodicity
shapes we are not aware of yet. In order to fully realize the
exploratory potential of those surveys, it is important to increase
the detection probability of the extreme cases. This is where the
techniques presented here become essential.

There are several new directions for further research that are now
open. Computationally, the calculation may be quite demanding, since
it involves rearranging the data values all over again for each period
(phase-folding). This operation, which is basically an operation of
sorting, is of complexity $O(N \log N)$. Repeating this for every
period may render the methods we introduced impractical for large
datasets. This warrants some algorithmic research to look for
fast techniques to perform the operation.

The very impressive results of the Hoeffding-test metric merit a
closer look at its definition and the theory behind it. As a test of
independence between two random variables, it is based on the measure
of deviation from independence of the two variables. Let us denote by
$G_1$ and $G_2$ the cumulative distribution functions of the two
variables, and by $G_{12}$ their joint cumulative distribution
function. Then, independence of the two variables would mean $G_{12} =
G_1 G_2$.  Armed with this definition, and using Cram\'er--von Mises
criterion for distance between distributions \citep{Cra1928,von1928},
Hoeffding defined $D$ by: 
$$
\int (G_{12} - G_1 G_2)^2\,dG_{12} \ ,
$$ 
where we use the empirical distribution function, determined only by
the observed values. This is somewhat reminiscent of the
Kolmogorov--Smirnov philosophy, which is popular among astronomers
\citep[e.g.][]{BabFei2006}.  This formula eventually results
in the formulae presented in Eqs \ref{HoeffD}--\ref{HoeffC}. 

\begin{figure}
\includegraphics[width=0.4\textwidth]{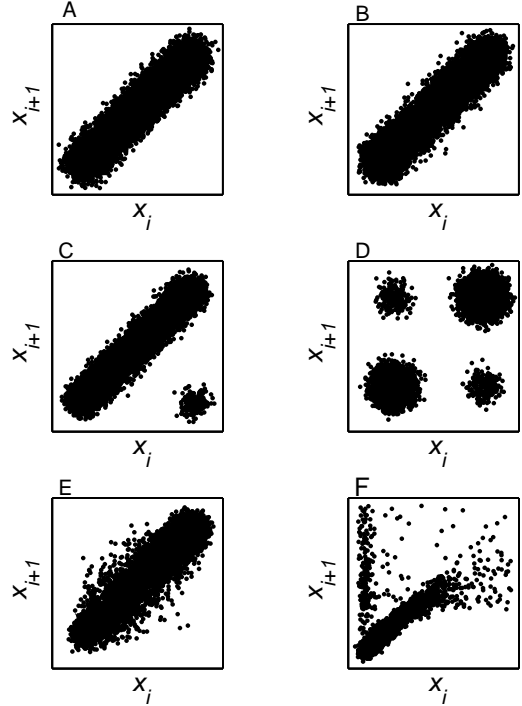}
\caption{The dependence between consecutive phase-folded samples for
each periodic-signal shape examined. The specific panels correspond to the
panels in Fig.\ \ref{shapes}: A. Sinusoidal; B. Almost sinusoidal;
C. Sawtooth; D. Pulse wave; E. Eclipsing binary light curve;
F. Eccentric spectroscopic binary RV curve.} 
\label{dependences}
\end{figure}

It is now clear why Hoeffding test tests for any kind of dependence,
not necessarily linear or even monotonous.  Fig.\ \ref{dependences}
provides further insight. For each periodic signal shape We simulated
$200$ light curves with $100$ points each, and with SNR $5$. We then
plotted the dependence between each sample and its successor, after
phase folding. In the simpler cases of sinusoidal and
almost-sinusoidal (panels A,B), the dependence seems completely
linear. This explains the satisfactory performance of von_Neumann
ratio in these cases. In the other panels the departure from linearity
may be quite significant, especially in the sawtooth and eccentric
binary RV cases. A strong departure from linearity is also apparent in
the pulse-shape periodicity, but one has to remember that the plot
seems very similar when folding in the wrong period, which explains
why the periodicity metric peak in those cases is not very prominent.

We also note here that we use the original expression that
Hoeffding introduced.  At a later stage, \citet*{Bluetal1961} proposed
an approximation to the Hoeffding statistic that is much easier to
compute, and in the context of periodicity detection it may open
possibilities for further simplifications. Hoeffding did not provide
any intuitive meaning to the various quantities used in the
calculation -- $A$, $B$, and $C$ (Eqs \ref{HoeffA}--\ref{HoeffC}). The
only one which seems to have an obvious meaning is $A$ (Eq.\
\ref{HoeffA}), which is a measure of the serial correlation of the
squared ranks.

In terms of the statistic used, recall that our rationale is based on
the rationale of the string-length methods. Besides pure serial
dependence measures, several authors tried to incorporate into the
string-length statistics also information related to the actual phase
of the measurements, penalizing pairs of consecutive points with large
phase difference \citep*[e.g.][]{Buretal1970,Ren1978}. This may be
attempted also in any one of the techniques we surveyed here, and
perhaps it may improve the performance even more.

In the simulations we present we use a completely random time
sampling, representing completely uneven sampling. In real-life
astronomical applications time sampling can have a quasi-periodic
nature, especially related to day-night alternation, but also to lunar
phase (monthly) and observability of the studied object (annual). This
is also true in a more complicated way for the cases of {\it Gaia} and
{\it Hipparcos}. This may somehow affect the simulation results,
probably by introducing aliases. However, since this paper is only the
first introduction of the new methods, we have not gone into the
details of exploring the full range of astronomical contexts.

Another idealization we have made in this work is considering only
white noise, with a prescribed SNR. Life is obviously more complex
than that, and in real-life signals there are also trends and
systematics ('red noise'). An obvious way to deal with those is by
applying a preliminary stage of detrending. Nevertheless, it would be
interesting to test the robustness of the non-parametric techniques
against such phenomena.  A similar problem is the problem of
multi-periodic signals, which are also a problem for the conventional
string-length methods. As real-life tests, one should also test the
techniques on existing sparsely sampled databases, e.g.,
\textit{Hipparcos} Epoch Photometry \citep{ESA1997}, and see whether
we detect all the known periodicities, and maybe add some more unknown
ones.

We focused here on tests of serial dependence of the phase-folded
data. However, there are many other tests of non-parametric
statistics, and there is potential for more innovative ways to turn
them into periodicity metrics.

While we introduce here five new kinds of periodicity metrics, it
seems that one of them -- the serial Hoeffding-test statistic --
emerges as a very promising new periodicity detection method,
that may improve periodicity detection significantly.  As part of its
'maturation' process, it should be further studied, both in terms of
the period detection efficiency and in terms of the estimation
accuracy, in various configurations. To promote further research and
testing of this periodicity metric we make it available online, in the
form of a Matlab function\footnote{The URL for downloading a Matlab
code to calculate the Hoeffding-test periodicity metric is
http://www.tau.ac.il/\~{}shayz/Hoeffding.m}.

\section*{acknowledgements}

This research was supported by the Israel Ministry of Science and
Technology, via grant 3-9082.

\end{document}